\documentclass[fleqn,10pt]{wlscirep}
\usepackage[utf8]{inputenc}
\usepackage[T1]{fontenc}
\usepackage[noabbrev, capitalise]{cleveref}
\usepackage{capt-of}
\usepackage{algorithm}
\newcommand{\amendRA}[1]{#1}
\newcommand{\amendRB}[1]{#1}

\title{Sublinear Domination and Core-periphery Networks}

\author[1,*]{Marios Papachristou}
\affil[1]{Cornell University, Computer Science Department, Ithaca, NY, 14850, United States.}

\affil[*]{papachristoumarios@cs.cornell.edu}
\newcommand{\ev}[2]{\mathbb E_{#1} \left [ #2 \right ] }


\begin{abstract}

In this paper we devise a generative random network model with core-periphery properties whose core nodes act as \emph{sublinear dominators}, that is, if the network has $n$ nodes, the core has size $o(n)$ and dominates the entire network. We show that instances generated by this model exhibit \emph{power law} degree distributions, and incorporates \emph{small--world} phenomena. We also fit our model in a variety of real--world networks. 

\end{abstract}

\begin{document}

\flushbottom
\maketitle

\thispagestyle{empty}

\maketitle


\section{Introduction}

Many complex networks exhibit the so--called \emph{core--periphery} structure. The core--periphery structure of networks considers a network that is comprised by a \emph{core} and a \emph{periphery}. The core of the network is a small subset of the node set which are \emph{tightly connected} with one another and the periphery ``lies around'' the core and nodes within the periphery are \emph{sparsely connected} with one another. In the context of a social network, the core of the network refers to the individuals that possess a \emph{celebrity status} in society, such as famous politicians, actors, and athletes, and the rest of the users constitute the periphery of the network.

Core-periphery networks have existed awhile in literature \cite{nemeth1985international, wallerstein1987world, zhang2015identification, snyder1979structural, zhang2015identification}. The intuition behind core-periphery networks has its roots in political economy. Wallerstein in his seminal work \emph{``World--systems theory''} \cite{wallerstein1987world} theorized that the globe can be divided into core nations, which focus on ``highly--skilled labor'' and ``capital--intensive'' production whereas peripheral countries focused on ``low--skilled labor'' and ``labor--intensive'' production. Moreover, trade and diplomatic ties between countries seem to follow this structure, backed by Krugman's theory \cite{krugman1991increasing} which argues that core-periphery structures emerge due to the core regions' low centralized-production costs and the supply-oriented peripheral regions. Avin et al. present an axiomatic approach towards core-periphery networks and draw strong conclusions  \cite{avin2014core}. Generative models for core--periphery networks have also been studied at~\cite{borgatti2000models, zhang2015identification, jia2019random, elliott2020core}. The closest model to ours is the stochastic blockmodel of~\cite{zhang2015identification} which assumes that \emph{core--core} nodes are connected with probability $p_{CC}$, \emph{periphery--periphery} nodes are connected with probability $p_{PP}$ and \emph{core--periphery} nodes are connected with probability $p_{CP}$, with $p_{CC} > p_{CP} > p_{PP}$, and its  recent extension to directed graphs in~\cite{elliott2020core}.

The \emph{dominating set} is a well--studied component of networks. More specifically, a subset of the nodes of an undirected network is a dominating set if and only if every node in the network has at least one neighbor belonging to the dominating set. The interesting question from an algorithmic perspective is finding the \emph{minimum dominating set} which is shown to be an \emph{NP--Hard} problem~\cite{garey1979computers}. Multiple previous works have investigated dominating sets in the context of social and biological  networks~\cite{bonato2015domination, molnar2013minimum, cooper2005lower, nacher2012dominating}. The work of~\cite{bonato2012geometric} shows that the \emph{geometric protean} model exhibits a sublinear dominating set, both in theory and in practice.

However, this previous modeling work used a generative framework that was quite complex, and lacked a connection with core-periphery structure. Here we present a much simpler generative model for networks whose minimum dominating set is sub-linear in size. We \emph{associate} the resulting minimum dominating set with the core of the network and its neighborhood (without including nodes of itself) to the periphery of the network. 
\amendRB{The main concept behind exploiting the core--periphery structure of networks to speed up computational tasks is based on the general idea that intense computational tasks can be performed within the \emph{sublinear core} and then the results can be aggregated to the periphery with relatively low query complexity. So, leveraging the connection between dominating sets and the core--periphery structure from an algorithmic viewpoint can be used in many problems such as all--pairs--shortest-paths computation, community detection, embedding generation, and many more.} 

We call our model the \emph{Influencer--Guided Attachment Model} (IGAM). The IGAM model is built onto  a \emph{hierarchical substructure}, also known as a \emph{communities--within--communities} (fractal--like) model~\cite{leskovec2007graph, kleinberg2002small}, that is a tree of \emph{fanout} $b$ and height $H$. Based on the tree skeleton, nodes are associated with \emph{prestige} (equivalently ``coreness'') values and between any two nodes, the log--probability of connection depends on the most prestigious node. The \emph{novelty} of IGAM concentrates on the existence of a \emph{sublinear minimum dominating set} which can be seen as defining the core of the network, with the rest of the nodes being the periphery of the network. We validate our hypothesis by efficiently fitting the IGAM model to real–
world data and show almost perfect correlation between
the construction of an almost dominating set based
on the IGAM model and the construction of an almost
dominating set via the maximum coverage greedy algorithm of \cite{nemhauser1978analysis}.  IGAM follows a \emph{power law} distribution, and exhibits \emph{small--world} phenomena, which are evident in social networks. 
\amendRB{
We compare the IGAM model with the logistic models of \cite{jia2019random, tudisco2019nonlinear} and conclude that IGAM is able to produce \emph{smaller} almost dominating sets than the logistic models of \cite{jia2019random, tudisco2019nonlinear}. 
}
\amendRA{
Finally, we give a generalized model that can incorporate core--periphery properties similar to the stochastic blockmodel of Zhang, Martin and Newman~\cite{zhang2015identification}
}.

\section{The Influencer--Guided Attachment Model} \label{sec:igam}


\amendRA{
Real--world networks usually exhibit \emph{power laws} together with \emph{self--similar artifacts}. Self--similar structures are  similar to a part of themselves and are common properties of \emph{fractals} \cite{schroeder1996fractals, leskovec2007graph}. Self--similar structures have been long observed in networks such as computer networks, patent networks, social networks \cite{watts2002identity,menczer2002growing,leskovec2010kronecker}. A model that is able to describe the communities--within--communities structure can be a tree structure. Moreover, we want a way to quantify that nodes have a higher affinity to be connected with more \emph{prestigious} nodes that are located higher in the tree rather than other nodes below their level, which refers to a common property of core--periphery networks \cite{zhang2015identification, borgatti2000models, wallerstein1987world}. This property is given by something which we call, similarly to \cite{leskovec2007graph}, a \emph{difficulty function}. We want the graph to follow a power law degree distribution as well as experience small--world phenomena \cite{kleinberg2002small}.

We are ready to describe the generative model formally: The model starts with a hierarchical structure of a perfect
\amendRA{$b$--ary tree}
$T$ of height $H$ and fanout $b \ge 2$ where $b$ is a constant. Every node $v$ of the tree is associated with a height $0 \le h(v) \le H$ which is defined to be the \emph{inverse prestige} of the corresponding node. The root has a higher prestige and as we go down on the tree the nodes have lower prestige up to the leaves. Two nodes $u$ and $v$ are linked with a probability equal to $f(u, v)$. We want $f(u, v)$ to depend on the node with the higher inverse prestige, and be scale-free. For the former property we can assume that $f(u, v)$ depends on $\min \{ h(u) , h(v) \}$. For $f$ to be scale--free we need $f$ to be level--independent (or translation--invariant). Namely, for two nodes $u, v$ at levels $h(u), h(v)$ and for two nodes $u', v'$ with $h(u') = h(u) - 1$ and $h(v') = h(v) - 1$ we must have that $f(u, v)$ and $f(u', v')$ to be level--independent and, thus, a constant multiplicative factor apart. Formally, if we let $\tilde h = \min \{ h(u), h(v) \}$ then $\min \{ h(u'), h(v') \} = \tilde h - 1$ that means $f(\tilde h) / f(\tilde h - 1) = c$, and subsequently, $f(\tilde h) \propto c^{-h}$ for some constant $c > 1$. This analysis yields a law of the form 

\begin{equation*} \tag{IGAM} \label{eq:igam}
    f(u, v) = c^{-1 - \min\{ h(u) , h(v) \}},
\end{equation*}
}
where $c \in (1, b)$ is a constant. The requirement that $c \in (1, b)$ will become evident as we go through this paper. After the generation of the random edges according to the law $f(u, v)$, we delete the auxiliary tree edges of $T$. For instance, if $b = 3$ and $c = 2$ then the root is connected with every leaf with probability $1/2$, the first level is connected with every leaf with probability $1/4$, and so on. The network has $n = \Theta(b^H)$ nodes. 
\amendRA{
\subsection{Basic Definitions} 

We say that a subset $S$ of the vertex set of a graph is a $\kappa$ almost dominating set ($\kappa$--ADS) if the set $S$ dominates at least $\kappa  n$ of the nodes present in the graph. In other words, at least $\kappa n$ of the nodes of the graph have a neighbor in $S$.

We say that an event $E(n)$ holds \emph{with high probability} (w.h.p.) if $\Pr [E(n)] = 1 - O(1 / n)$, \emph{with extreme probability} (w.e.p.) if $\Pr [E(n)] = 1 - O(e^{-n})$, and \emph{asymptotically almost surely} (a.a.s.) if $\Pr [E(n)] \to 1$ as $n \to \infty$. 
 }

\section{Sublinear Domination and Core--periphery Structure} \label{sec:dominating_set}

We show that the \emph{core} of the network consists, as one should expect, from a \emph{sublinear} number of nodes located at the top levels of the tree. To observe this phenomenon, we calculate the probability $q_{h \tau}$ of a node at height $h$ not being dominated by any node between levels $0$ and $\tau$ where $\tau \le h$, which equals 

\amendRA{
\begin{equation*}
    q_{h \tau} = \prod_{r = 0}^\tau \left ( 1 - c^{-r - 1} \right )^{b^r} \le e^{- \frac 1 c \sum_{r = 0}^\tau (b / c)^r } \lesssim e^{- \frac 1 c \left ( \frac b c \right )^{\tau + 1}} = e^{- \Theta \left ( \left ( \frac b c \right )^\tau \right )},
\end{equation*}
}

\amendRA{
where $a \lesssim b$ denotes that there exists a constant $C > 0$ independent of $b$ such that $a \le C \cdot b$ (i.e. inequality up to a constant factor),
}
and the first inequality holds since $1 - t \le e^{-t}$ for all $t \in \mathbb R$. Note that $q_{h \tau}$ does not depend on the height of the node in question, as long as $\tau \le h$. Now the probability that there is at least one node uncovered below level $\tau + 1$ is given by Markov's Inequality and is at most 

\amendRA{
\begin{equation*}
    \sum_{h = \tau + 1}^H b^h q_{h \tau} = q_{h \tau} \sum_{h = \tau + 1}^H b^h \lesssim q_{h \tau} b^H \lesssim e^{H \log b - \Theta \left ( \left ( \frac b c \right )^\tau \right )}. 
\end{equation*}
}

To assert an w.h.p. guarantee we force the above probability to be $\Theta (b^{-H})$, therefore, solving for $\tau$ we arrive at a dominating set of size 

\begin{equation*}
    n_0 = b^{O(\log (2 cH \log b) / \log (b / c))} = b^{o(H)} = o(n)
\end{equation*}

with probability at least $1 - \Theta(b^{-H})$. Consequently, a \emph{sublinear} fraction of nodes $C = \{ v : h(v) \le \tau \}$ located on a \emph{logarithmic height} $\tau$ from the root of the skeleton tree $T$ dominate the whole periphery $P = \{ v : h(v) \ge \tau + 1 \}$ with $\tau = \Omega (\log H)$. 



\section{Degree Distribution} 

To fit the model to real--world data, we infer the degree distribution of IGAM. The average degree of a node $u$ at level $h$ is 
\begin{equation*}
\begin{split}
    \bar d_h & \approx \sum_{r = 0}^H b^r c^{-\min \{ h, r \} - 1} \\
    & =  \frac 1 c \left [ \left ( \frac b c \right )^{h + 1} - 1 + \frac {b^{H + 1} - b^{h + 1}} {c^h} \right ] = \Theta \left ( \frac {b^{H + 1}} {c^{h + 1}} \right )  . 
\end{split}
\end{equation*}

and the total expected number of edges at level $h$ is $\bar m_h = b^h \bar d_h = \Theta (b^{h + H + 1} / c^{h + 1})$. The asymptotics of the previous Equation yield a power law with exponent

\begin{equation*}
    \frac {\mathrm{d} \log \bar d_h} {\mathrm d h} = \log \left ( \frac 1 c \right ).
\end{equation*}



If the rank of $u$, with $h(u) = h$ is given as $r_h = c^{h}$, which is an increasing function of $h$, then the expected degree depends on the inverse rank $1 / r_h$, yielding a \emph{Zipfian} power law. The trials for connecting every node are independent Bernoulli variables, and therefore by the multiplicative Chernoff bound with probability at least $1 - \Theta (b^{- H})$ we have that the average degree at height $h$, $\hat d_h$ is $\Theta(1 / r_h)  \pm O(\sqrt {H \log b / (2b^h)})$. 
\amendRA{
By a union bound, we have that 

\begin{equation*}
    \Pr \left [\exists h \in \{ 0, \dots, H \} : \left |\hat d_h - \Theta \left ( \frac 1 {r_h} \right ) \right | = \Omega \left ( \sqrt {\frac {H \log b} {2b^h}} \right ) \right ] \le \sum_{h = 0}^H \Pr \left [ \left |\hat d_h - \Theta \left ( \frac 1 {r_h} \right ) \right | = \Omega \left ( \sqrt {\frac {H \log b} {2b^h}} \right ) \right ] = O \left ( \frac {H} {b^H} \right ). 
\end{equation*}

Thus, with probability $1 - O(H b^{-H})$ (i.e. w.h.p.) the degree histogram follows Zipf's Law. 
}
The exponent of the degree distribution can be altered, if the same model is generated with parameters $b' = b^\alpha, c' = c^\alpha$ for some $\alpha \ge 0$. The expected number of edges $\bar m$ is given as 

\begin{equation*}
    \bar m = \frac 1 2 \sum_{h = 0}^H b^h \bar d_h = \Theta \left ( \frac {b^{2H}} {c^H} \right ),
\end{equation*}

and is \emph{superlinear} with respect to the number of nodes.

\section{Insights from Data} \label{sec:igam_fit}

We  describe a fitting algorithm for the IGAM model (Algorithm \ref{alg:igam_fit}). The fitting process considers of being given a sample of $m$ edges $\mathcal D = \{ e_i \}_{1 \le i \le m}$ on a network of $n$ nodes where $n$ is known. Our goal is to find the optimal fanout $ b^\star$, the optimal height function $h^\star$ and the optimal scale factor $c^\star$ that maximize the log--likelihood of the model, that is 

\begin{equation*}
    \max_{b, \; h, \; c} \ell (b, h, c | \mathcal D) = \max_{b,\; h, \; c} \log \Pr [\mathcal D | b, h, c].
\end{equation*}

\amendRB{
where the likelihood equals 

\begin{equation*}
    \ell (b, h, c | \mathcal D) = \sum_{(u, v) \in E} \log f(u, v) + \sum_{(u, v) \notin E} \log (1 - f(u, v)) = \sum_{(u, v) \in E} \log \left ( \frac {f(u, v)} {1 - f(u, v)} \right ) + \sum_{u, v \in V \times V} \log (1 - f(u, v)).   
\end{equation*}
}
Directly optimizing the likelihood is very hard since there are $O(n)$ possible fanouts, each fanout can generate an exponential number of possible trees, and thus height functions, and given the fanout and the height function the remaining problem consists of finding the optimal $c$ that explains $\ell (c | \mathcal D, b, h)$. 

To optimize the log--likelihood of IGAM efficiently, we first calculate the sample degrees of each node, that is $\bar y_u = \sum_{i = 1}^m \mathbf 1 \{ u \in e_i \}$, and then order the nodes on decreasing order of their sample degrees. After that, we fix a fanout $b$ from the interval $\{ 2, \dots, n - 1 \}$, and according to that fanout we start by attributing heights of a hypothetical $b$--ary tree on the nodes according to their descending order. For example, for $b = 2$ the first node gets a height of 0, the next two a height of 1, and so on. Then, for each height $0 \le h \le \lceil \log n / \log b \rceil$, we form the log--degrees $\bar z_h = \log \left ( \sum_{u : h(u) = h} \bar y_u  \right )$, and fit linear--least--squares with $x$--values being the range of heights and $y$--values being the log--degrees $\bar z_h$. The optimal slope $a$ yields $c$ to be $c = b \cdot e^{-a}$. If $c \ge b$ then the current fit is rejected. We can then calculate the likelihood function $\ell$ and keep the best parameters $(b^\star, h^\star, c^\star)$. 
\amendRB{
Each step is dominated by the calculation of the likelihood that costs $O(n^2)$ time, since \emph{exactly} computing the log--likelihood requires summing over all pairs of nodes (regardless of whether an edge exists or not),  and thus the total complexity is $O(n^3)$. Note that since the values of $f(u, v)$ are small (i.e. close to 0) and real--world networks are sparse (i.e. $m$ is of the order of $n$ or $n \log n$) the log--likelihood can be \emph{approximated} in time $O(m)$ by ignoring the network--independent term, i.e. the term that sums on $V \times V$, which yields an algorithm with runtime $O(nm)$ instead of $O(n^3)$. 
}
If a full  $b$--ary tree does not cover the network, we allow the last level to be incomplete.
  
\begin{figure}[ht]
    \centering
    \includegraphics[width=\linewidth]{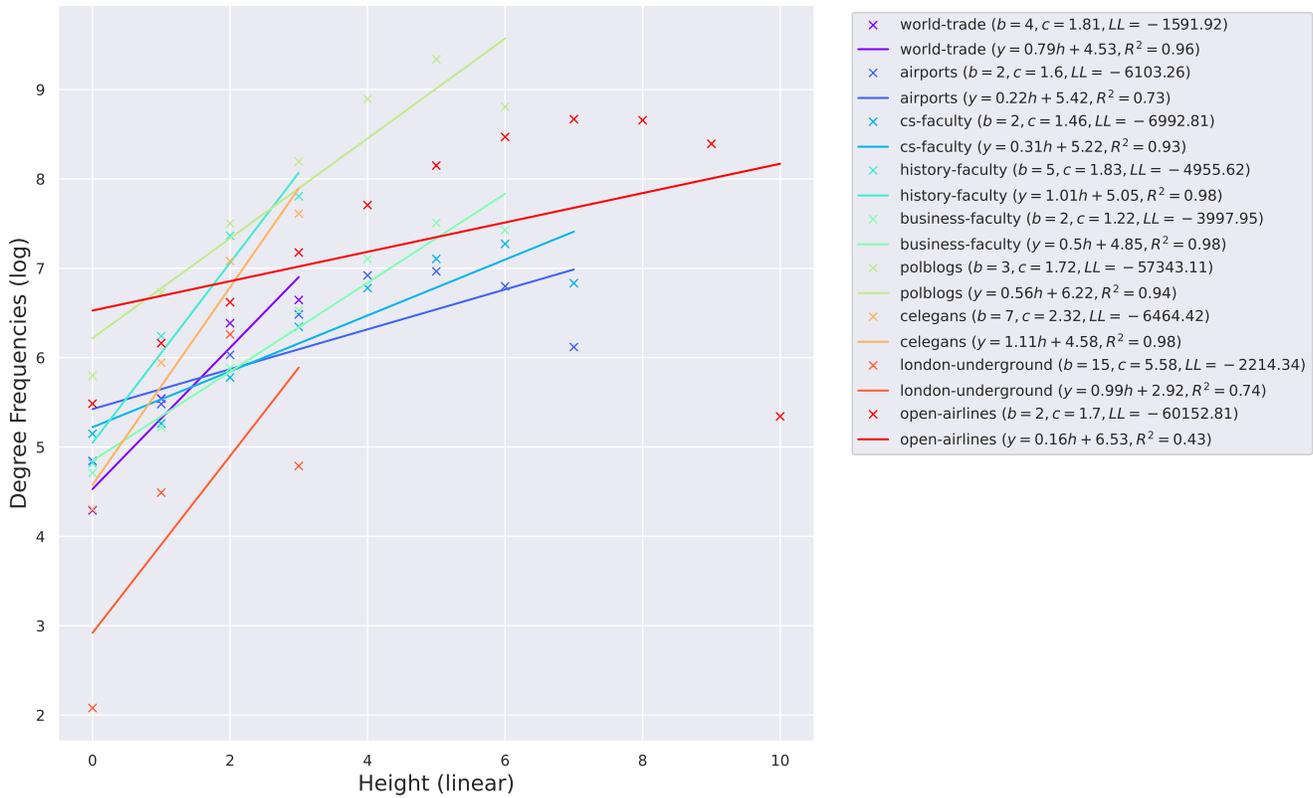}
     
    \caption{Results of fitting an IGAM model to the world--trade, cs--faculty, history--faculty, business--faculty, and airports datasets examined in~\cite{elliott2020core, de2018exploratory, clauset2015systematic, colizza2007reaction, adamic2005political}. The Figure displays the predicted values of $b$ and $c$ for the IGAM model, and the total degree at each level $h$ of the skeleton tree of fanout $b$. A linear fit is presented for each dataset to showcase the power law behaviour. Moreover values of the log--likelihood (LL), and Pearson's Correlation Coefficient $R^2$ are reported. Nodes with degree $\le 4$ have been filtered out as outliers except for the london--underground network. Source code to reproduce the Figure can be found in\cite{source_code}.}
    \label{fig:igam_fit}
\end{figure}

\begin{figure}[ht]
    \centering
    \includegraphics[width=\linewidth]{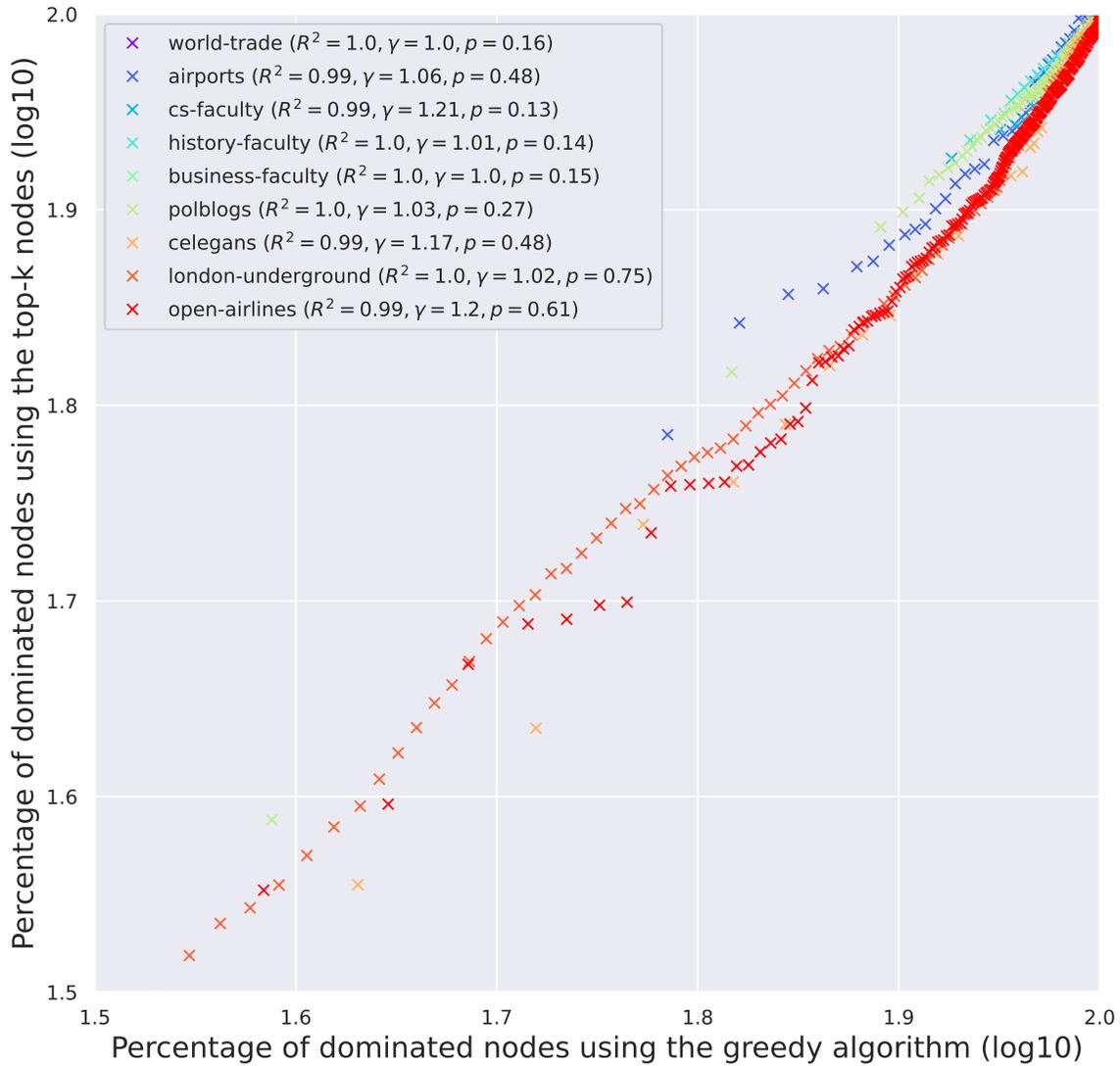}
    \caption{Log--log plot between the percentages of dominated nodes when running the greedy $(1 - 1 / e)$--maximum coverage algorithm of~\cite{nemhauser1978analysis} ($x$--axis) and selecting nodes according to their hierarchy, i.e. in order of descending initial degree ($y$--axis). The slope $\gamma$ and $R^2$  of linear fits are reported. 
    \amendRA{
    The rule that selects nodes based on their prestige $h$ yields very close results to the greedy maximum coverage algorithm. In general instances, these two algorithms are expected to have different results, since the former algorithm may select prestigious nodes whose neighborhoods have large overlaps which may not yield good coverage in general. However, specifically in core--periphery networks, high prestige nodes seem to have small overlaps, which justifies the good performance of the prestige--based algorithm. Source code to reproduce the Figure can be found in\cite{source_code}.
    }}
    \label{fig:dominating_set}
\end{figure}

\begin{figure}[ht]
    \centering
    \includegraphics[width=0.75\linewidth]{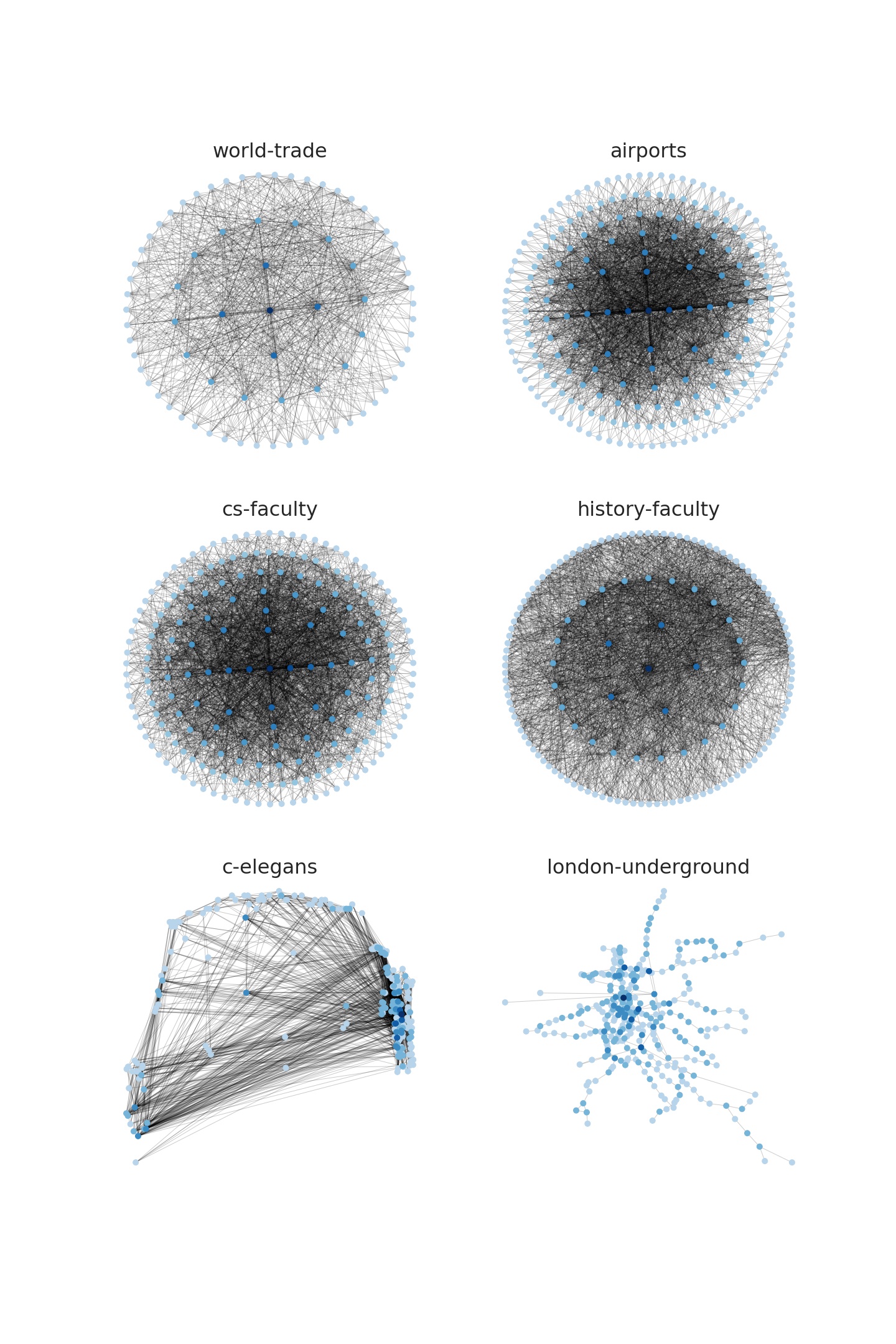}
    \caption{IGAM model fitted on small datasets (world--trade, airports, cs--faculty, history--faculty, c--elegans, london--underground). The darker colors refer to nodes with higher prestige and the lighter colors refer to nodes with lower prestige. Source code to reproduce the Figure can be found in\cite{source_code}.}
    \label{fig:visualization_igam_fit}
\end{figure}

We fit the IGAM model to networks examined in~\cite{elliott2020core}. More specifically, we examine the world--trade network from~\cite{de2018exploratory} ($n = 76, \; m = 845$), the faculty datasets from~\cite{clauset2015systematic} (cs--faculty: $n = 205, \; m = 2,861$; history--faculty: $n = 145, \; m = 2,334$; business--faculty: $n = 113, \; m = 3,027$), the polblogs dataset from~\cite{adamic2005political} ($n = 852, \; m = 15,956$), the airports dataset from~\cite{colizza2007reaction} 
\amendRB{($n = 210, \; m = 2,429)$, the c--elegans dataset from \cite{Kaiser-2006-placement}  ($n = 279, \; m = 1.9 \mathrm K$), the open--airlines dataset  from~\cite{jia2019random} ($n = 7.2 \mathrm K, \; m = 18.6 \mathrm K$), and the london--underground dataset from \cite{jia2019random} ($n = 315, \; m = 270$);}
treating the networks as undirected. Figure~\ref{fig:igam_fit} presents the (total) degree distribution fits for the IGAM model, where the parameters $b, c$ and the height function have been determined. Observe, that the total degree at each constructed level is linearly correlated ($R^2 \ge 0.93$ except for the airports dataset) with the coreness value of each group of nodes (per level). Moreover, in Figure~\ref{fig:dominating_set} we do a log--log plot between the construction of the dominating set as in Section~\ref{sec:dominating_set} and the construction of the dominating set using the maximum coverage greedy algorithm. 
\amendRB{
    The former algorithm treats the nodes as IGAM would do in the construction of the dominating set, i.e. by traversing the levels of the hierarchy from top to bottom. The latter algorithm picks the node with the largest \emph{active degree} at each step, adds it to the set, and removes itself and all the nodes connected to it from the network up to a certain number of steps or if there are no more nodes left. 
}
Markers on the plot represent subsequent iterations of both algorithms. We observe almost perfect correlation between the two algorithms and slightly superlinear relations of the form $y \propto x^\gamma$ for $\gamma \in [1, 1.21]$, which is a phenomenon that we should not expect in more general networks, since choosing the nodes with the highest degrees shall not yield good coverage in general. Moreover, note that a sublinear number of iterations, denoted by the number of \texttt{x} markers outside the $[1.9, 2.0]^2$ box (the mapping is increasing),  suffices to dominate $10^{1.9} \% \approx 80 \%$ of the nodes. A visualization of the IGAM fitting process can be found in \cref{fig:visualization_igam_fit} for the small datasets whereas the various levels $h$ of the IGAM model are color--coded.

\subsection{Qualitative Insights} \label{sec:qualitative}

\amendRA{
In this Section, we highlight the following structures that emerge from fitting the IGAM model to the real world data. We examine the first three levels of the hierarchy, devised by the height function $h$, for the datasets that contain labeled nodes. The analytical form of the core nodes can be found in the Methods Section. 

\begin{enumerate}
    \item \emph{Faculty Networks.} In the faculty networks of each one of the three disciplines (computer science, history, and business) the core consisted of nodes referring to highly ranked universities in the United States (in each discipline), as well as an (aggregate) node referring to faculty coming from all non--US academic institutions. To elaborate, the cs--faculty network contains MIT, CMU, Stanford, UT Austin, Purdue, and UIUC in its core, together with the aggregate node. The history--faculty core consists, for instance, of Harvard, Yale, University of Chicago, Columbia, Stanford, Johns Hopkins, and Cornell. Finally, the business--faculty network has, for instance, the University of Michigan, UT Austin, Penn State, and the University of Pennsylvania at its core. These findings are consistent with the body of research on faculty hiring networks\cite{clauset2015systematic, lee2021dynamics} where it is stated that, for the computer science discipline, a very small percentage (9\%) of departments is responsible for 50\% of academic hires in faculty position. 

    \item \emph{Open--airlines.} The open--airlines network has a core that consists of very large and central international airports such as AMS, FRA, CDG, IST, MUC, ATL, and PEK.  
    
    \item \emph{World--trade.} The world--trade dataset contains data about the trade of metals among 80 countries in 1994. The nodes represent countries who have available entries in the Commodity Trade Statistics released by the United Nations. In this network the core consists of, for instance,  from Finland, Hungary, Slovenia, Singapore, Chile, and so on.   
    
    \item \emph{London--underground.} In the london--underground dataset, we recover a core that consists of busy train stations such as Bank, Baker Street, Canning Town, and so on, all of which are cardinal to the British underground system.
    
\end{enumerate}
}

\subsection{Relation to Logistic Core--periphery Models} \label{sec:comparison}

\begin{figure}[ht]
    \centering
    \includegraphics[width=\linewidth]{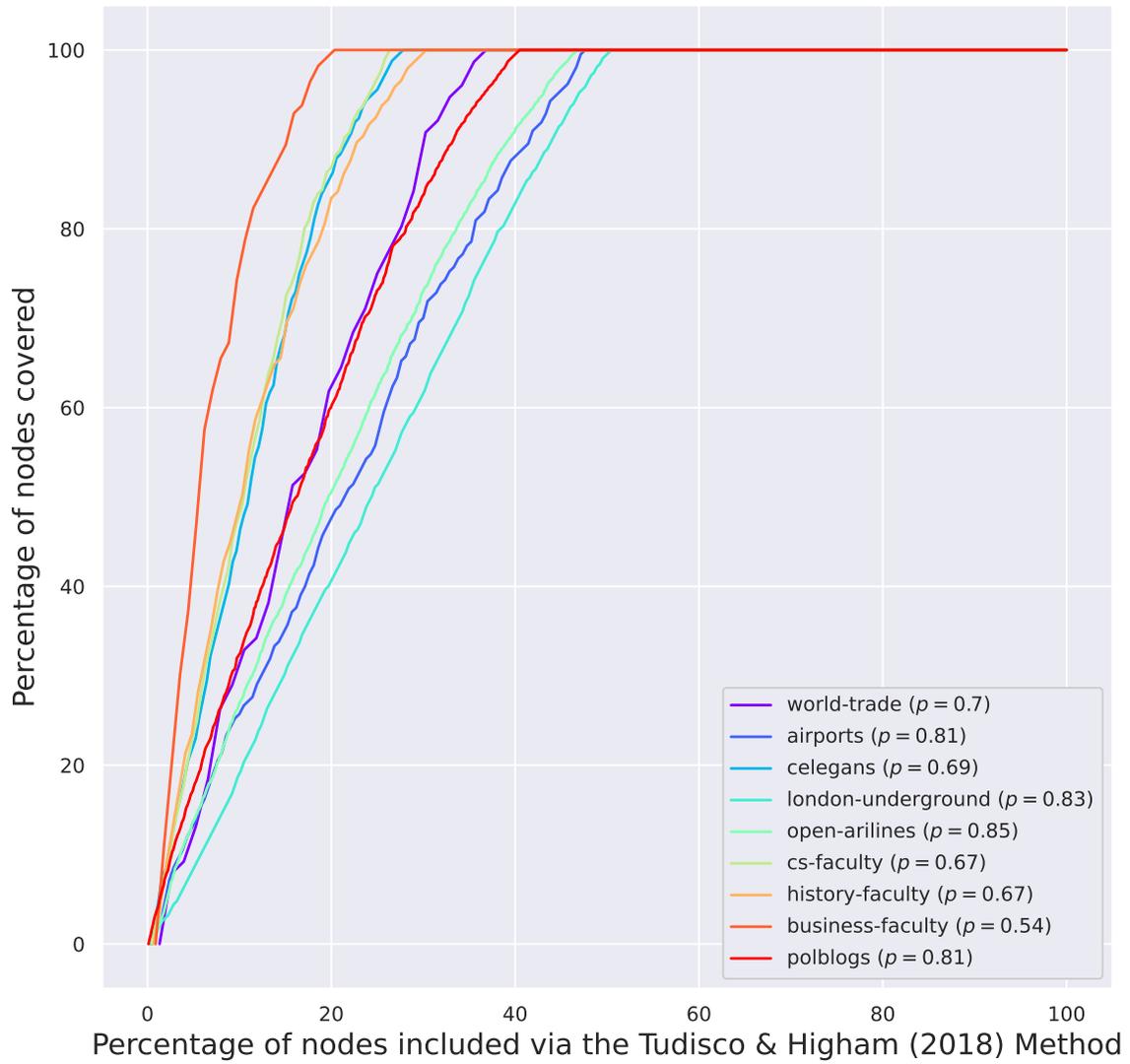} 
    \caption{Domination Curve by running the method of Tudisco and Higham \cite{tudisco2019nonlinear}. Source code to reproduce the Figure can be found in\cite{source_code}.}
    \label{fig:tudisco_higham}
\end{figure}

\begin{figure}[ht]
    \centering
    \includegraphics[width=\linewidth]{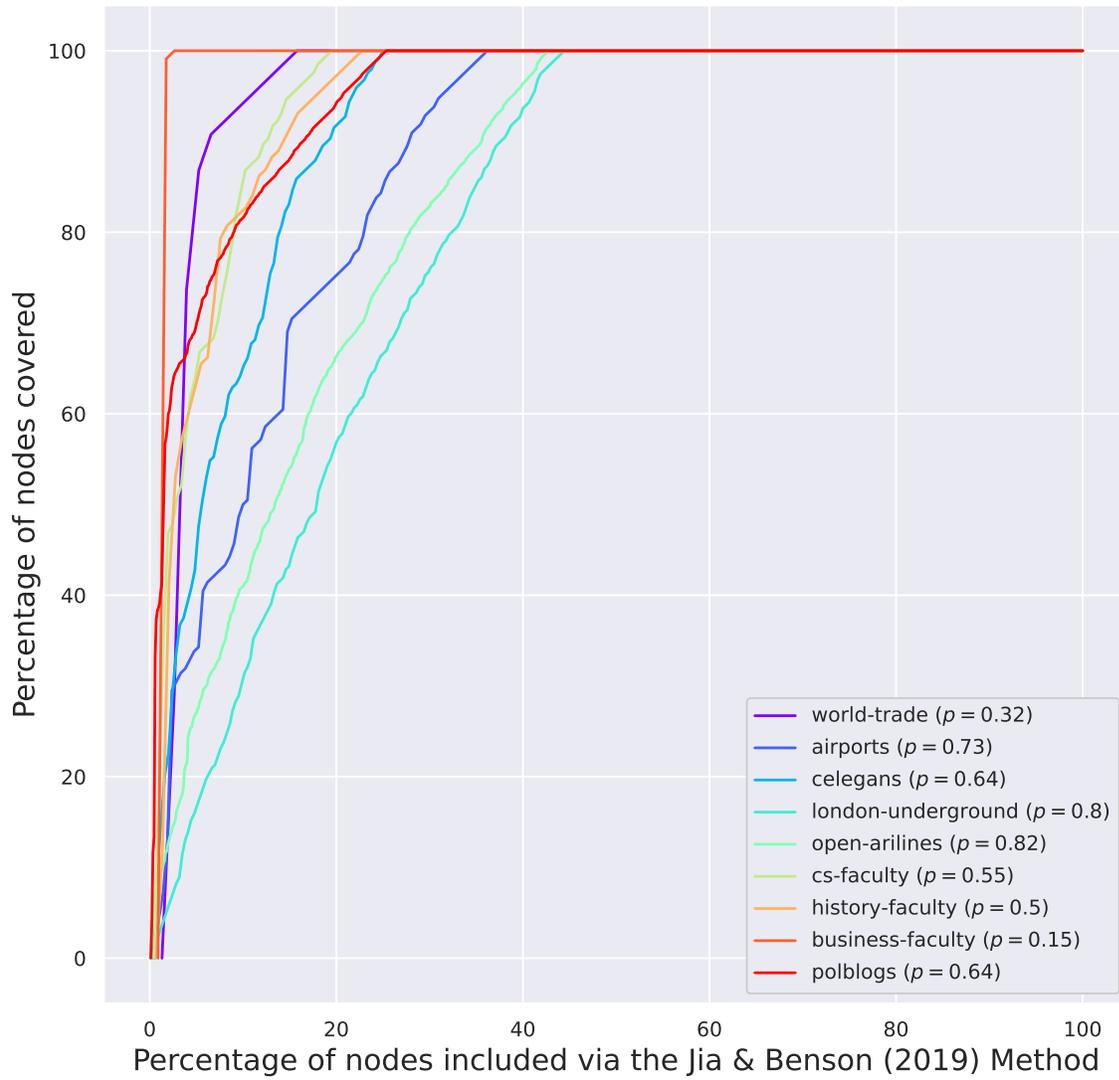}
    \caption{Domination Curve by fitting the model of Jia and Benson\cite{jia2019random} on spatial data and the logistic CP model otherwise. Source code to reproduce the Figure can be found in\cite{source_code}.}
    \label{fig:jia_benson}
\end{figure}

\begin{figure}[ht]
    \centering
    \includegraphics[width=0.65\linewidth]{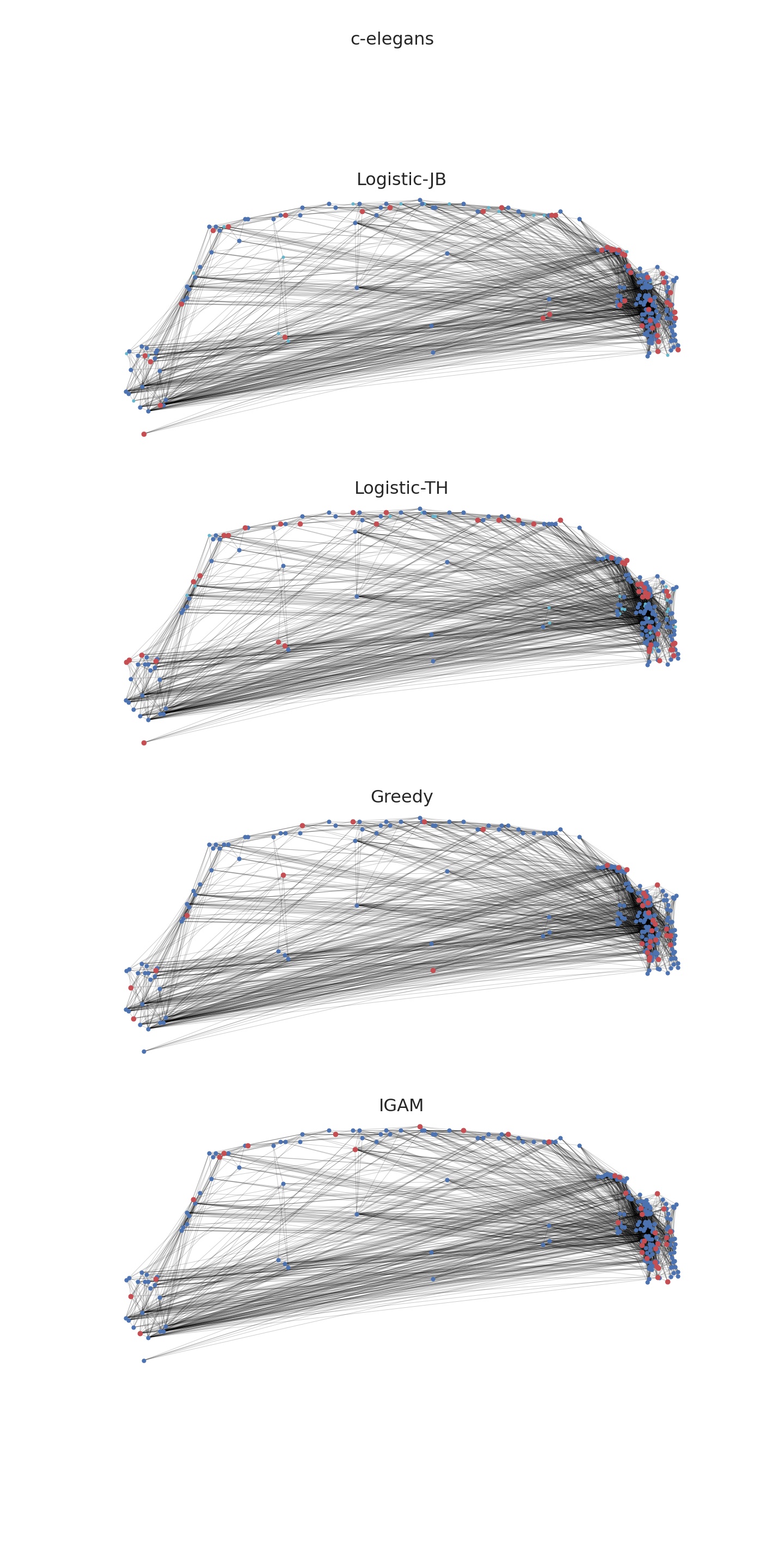}
    \caption{Visualization of a core set of size $n^{0.7}$ for the Logistic--JB, Logistic--TH, Greedy, and IGAM strategies. The red nodes represent members of the core set, the blue nodes are dominated nodes, and the cyan nodes are non--dominated nodes. Source code to reproduce the Figure can be found in\cite{source_code}.}
    \label{fig:visualization}
\end{figure}

\amendRB{
We compare the IGAM model with two logistic models introduced by Jia and Benson\cite{jia2019random} and Tudisco and Higham\cite{tudisco2019nonlinear}. In detail, we fit both models and give empirical answers to the following question: \emph{Are the logistic core-periphery models able to explain the domination structure of core--periphery networks?}

The model of Jia and Benson assigns a \emph{coreness score} $\theta_v \in \mathbb R$ for every node $v$ in the vertex set $V$. The simple version of the model produces edges $(u, v)$ randomly and independently with probability 

\begin{equation*} \tag{Logistic--CP} \label{eq:logistic_cp}
    \rho(u, v) = \frac {1} {1 + e^{-\theta_u - \theta_v}}.
\end{equation*}

Intuitively what this model describes is that a node with $\theta_v \ge 0$ is considered to be a core node and a node with $\theta_v < 0$ to be a peripheral node. That is, for a pair $(u, v)$ if both nodes are peripheral, i.e. have $\theta_u < 0$ and $\theta_v < 0$, then the link probability $\rho(u, v)$ is less than the case when one of $\theta_u, \theta_v$ is non--negative that represents a core--periphery link. Similarly, when both $\theta_u \ge 0$ and $\theta_v \ge 0$, which corresponds to a core--core node, then the edge creation law attributes a larger connection probability. When spatial features $x: V \to \mathbb R^d$ are provided, as well as a kernel function $K(u, v)$ (for example, $K(u, v) = \| x_u - x_v \|_2$), and a hyperparameter $\varepsilon$, then \eqref{eq:logistic_jb} is generalized to an edge law 

\begin{equation*} \tag{Logistic--JB} \label{eq:logistic_jb}
    \mu(u, v) = \frac {e^{\theta_u + \theta_v}} {K^{\varepsilon}(u, v) + e^{\theta_u + \theta_v}}.
\end{equation*}

The model of Tudisco and Higham\cite{tudisco2019nonlinear} is based on a logistic probability law determined by a ranking $\pi$ of the nodes. The more prestigious a node $v$ is the higher the value $\pi_v$ is. The edge creation law is given by

\begin{equation*} \tag{Logistic--TH} \label{eq:logistic_th}
    \varphi(u, v) = \sigma_{s, t} \left ( \frac {\max \{ \pi_u, \pi_v \}} {n} \right ),
\end{equation*}

where $\sigma_{s, t} = 1 / (1 + e^{-s (x - t)})$ is the smooth approximation of the Heaviside step function $H_t(x)$ that is 1 if $x \ge t$ and 0 otherwise. We use $s = 10$, and $t = 1/2$. Again, the model intuitively says that nodes tend to be associated with more prestigious nodes rather with less prestigious nodes. Finally, the authors propose an iterative method to infer the ranking $\pi$ which has an $O(m)$ per-step cost. 



We evaluate how well can Logistic--CP, Logistic--JB and Logistic--TH capture the domination properties of the core--periphery structure compared to IGAM. For the logistic models of Jia and Benson we fit the Logistic--CP model when there are no spatial data available and the Logistic--JB when spatial data are available (i.e. in the c--elegans, open--airlines, and london--underground datasets). We use the optimal parameters $\theta_v^*$ of the logistic models to build a ranking for the nodes by sorting them in decreasing order of the scores $\theta_v^*$. For the Logistic--TH model, we use the iterative method provided in their paper to infer the ranking by sorting the entries of the fixed point that their algorithm produces. Then, for all models, we report the domination curves in \cref{fig:dominating_set,fig:jia_benson,fig:tudisco_higham}. To give better visual insights on how the models perform, we visualize the outcome of fitting the models for the c--elegans dataset on \cref{fig:visualization} for a core set of size $\lfloor n^{0.7} \rfloor$. For each dataset and figure we report the exponent $p \in [0, 1]$ of a set that dominates 80\%  of the network (i.e. and 0.8--ADS). Namely, if a fraction $\varpi \in [0, 1]$ suffices to cover at least 80\% of the network, then $p = \log (\varpi \cdot n) / \log n$.  

\subsubsection{Key Takeaway} \label{sec:takeaway}
 
\amendRB{   
The IGAM model can better explain the sublinear domination phenomenon in core--periphery networks than Logistic--JB, Logistic--CP, and Logistic--TH. Also Logistic--CP and Logistic--JB achieve better coverage compared to Logistic--TH. Perhaps the most characteristic are the faculty (cs--faculty, history--faculty, business--faculty) and the world--trade datasets where IGAM produces an almost  dominating set with an exponent $p \le 0.16$ whereas Logistic--TH finds a similar set with $p \ge 0.54$, and Logistic--CP finds an 0.8--ADS with $p = 0.15$ in the case of business--faculty and with $p \ge 0.32$ in the rest of the datasets. In the polblogs dataset, IGAM is able to find an 0.8--ADS with $p = 0.27$ whereas Logistic--CP finds one with $p = 0.64$ and Logistic--TH finds a much larger one with $p = 0.81$. In the open--airlines dataset the 0.8--ADS corresponds to $p = 0.61$ for IGAM and to $p \ge 0.82$ for the logistic methods. Finally, the smallest variation between the methods exhibits the london--underground dataset where $p$ ranges from $p = 0.75$ (IGAM) to $p = 0.85$ (Logistic--TH). Concluding, the ADS constructed by IGAM are consistently smaller than the ones produced by Logistic--CP and Logistic--JB which are smaller than the ones produced with Logistic--TH, which suggests that IGAM is able to \emph{explain} the sublinear domination phenomenon where other logistic models \emph{fail} to do so.
}
}

\section{Miscellaneous Properties}

\subsection{Small--world Behaviour} To determine the diameter (the diameter of a disconnected network is taken to be the diameter of its giant connected component) of the network, we build an Erd\"os-Renyi (ER) network \amendRB{$W$} with $n$ nodes and edge probability $f^* = \min_{u, v} f(u, v) = c^{-H-1}$. It follows from a standard coupling argument, i.e. a ``toss--by--toss'' comparison, that we can relate the two networks as one being subgraph of the other, in our case the ER network \amendRB{$W$} being subgraph of the IGAM network, say $G$. The coupling is constructed as follows: $\Pr \left [(u, v) \in E(G) \; | \; (u, v) \in E(W) \right ] = 1$, $\Pr \left [(u, v) \in E(G) \; | \; (u, v) \notin E(W) \right ] = \tfrac {f(u, v) - f^*} {1 - f^*}$, so that $\Pr \left [(u, v) \in E(G) \right ] = f(u, v)$, and $\Pr [(u, v) \in E(W)] = f^*$. Then it follows that the diameter of the IGAM network is at most the diameter of \amendRB{$W$}. Using a result from~\cite{randomgraphs, chung2001diameter}, we have that since the average degree of \amendRB{$W$} is $\Theta((b / c)^H) \to \infty$ as $H \to \infty$, the diameter of \amendRB{$W$} is close to  $\log n / \log (n f^*) = \Theta (\log b / \log (b / c)) = O(1)$ a.a.s. From that we can deduce that $G$ has a diameter close to $O(\log b / \log(b / c)) = O(1)$ a.a.s. This result can follow from intuition also, since all the nodes at a logarithmic height of the root dominate the periphery and a worst case path should roughly be between two peripheral nodes which are connected via a node at the core, with this node being a common dominator of them. 
\amendRA{
\subsection{Global Clustering Coefficient (GCC)} The probability of $uvw$ being a triangle given that $h(u) \le h(v) \le h(w)$ is $\beta_{uvw} = f(u, v) f(u, w) f(v, w) = c^{-3 - 2 h(u) - h(v)}$, thus the expected total number of closed triangles $T_C$ is 

\begin{equation*}
    \mathbb E [T_C] = \sum_{(u, v, w): h(u) < h(v) < h(w)}b^{h(u) + h(v) + h(w)} \beta_{uvw} = \Theta \left ( \frac {b^{3H}} {c^{3H + 3}} \right ).
\end{equation*}

The calculation has been deferred to the Methods Section (\cref{eq:closed_triplets}). The probability $\gamma_{uvw}$ of $uvw$ being a triplet (open or closed) is given as $\gamma_{uvw} = f(u, v) f(u, w) + f(u, v) f(v, w) + f(u, w) f(v, w)$. Conditioned on the event that $h(u) \le h(v) \le h(w)$ we can deduce that $3 c^{-2-2h(v)} \le \gamma_{uvw} \le 3 c^{-2-2h(u)}$. Similarly to $T_C$, the expected number of open triplets $T_R$ is $\Theta \left ( \frac {b^{3H}} {c^{2H + 2}} \right )$ (see \cref{eq:open_triplets} in the Methods Section). By McDiarmid's Inequality~\cite{doob1940regularity}, since $T_C$ and $T_R$ are $\Theta(b^H)$-Lipschitz we have that $\Pr \left [ | T_C - \ev {} {T_C} | = \Omega (b^{H}) \right ]  = O \left (e^{-b^H} \right )$, and $ \Pr \left [ | T_R - \ev {} {T_R} | = \Omega (b^{H}) \right ]  = O \left (e^{-b^H} \right )$ and therefore we can deduce that the GCC $T_C / T_R$ is $O(c^{-H} + b^{-H}) = O(c^{-H})$ with probability $1 - O \left (e^{-b^H} \right )$ by combining the two concentration bounds. Therefore, w.e.p. clustering coefficient is $O(c^{-H})$. 

\subsection{Core--periphery Conductance} The expected conductance of a set $\emptyset \subset S \subset [n]$ is given as $\bar \phi(S) = \ev {} {e(S, \bar S)} /  \min \{ |S|, |\bar S| \}$, 
where $\bar S = [n] \setminus S$. Letting $S_\tau$ to be the nodes at the first $\tau < H$ levels where $|S_\tau| \le |\bar S_\tau|$ yields $\min \{ | S_\tau | ,  | \bar S_\tau | \} = b^{\tau + 1} - 1$, and 

\begin{equation*}
    \ev {} {e(S_\tau, \bar S_\tau)} = \sum_{s = \tau + 1}^H \sum_{r = 0}^\tau b^r b^s c^{-1- \min \{r, s \}} = \frac 1 c \sum_{s = \tau + 1}^H b^s \sum_{r = 0}^\tau \left ( \frac b c \right )^r = \frac 1 c \sum_{s = \tau + 1}^H b^s \Theta \left ( \left ( \frac b c \right )^\tau \right ) = \Theta \left ( |\bar S_\tau|  \left ( \frac b c \right )^\tau \right ). 
\end{equation*}

$\bar \phi(S_\tau) = \Theta ( b^H / c^\tau)$. Letting $\tau = \log (2 c H \log b) / \log (b / c)$ be the core's height we deduce that $\bar \phi(C) = \Theta ( b^H / H )$. 
}

\section{Model Generalizations} \label{sec:model_generalization}

\subsection{IGAM2}

\begin{figure}[ht]
    \centering
    \includegraphics[width=\linewidth]{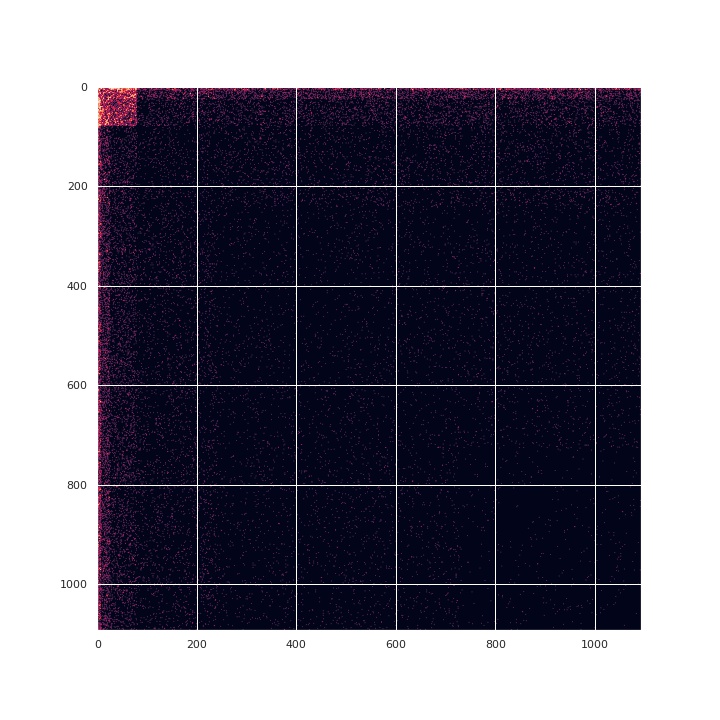} 
    \caption{Adjacency matrix of IGAM2 model with $c_1 = 1.5, c_2 = 2.5, b = 3$, $H_0 = 2$ and $H = 6$.}
    \label{fig:igam2}
\end{figure}

We fully align with the stochastic blockmodel definition of core--periphery networks presented in~\cite{zhang2015identification} by defining the following generalization of IGAM, which we call IGAM2, parametrized by $b > c_2 > c_1 > 1$. In this context, we start with the same skeleton tree of fanout $b$ and then the law $g(u, v)$ for generating the edges is 

\begin{equation*} \tag{IGAM2} \label{eq:igam2} 
    g(u, v) = \begin{cases}
        c_2^{-1-\min \{h(u), h(v) \}} & \max \{ h(u), h(v) \} > H_0 \\
        c_1^{-1-\min \{ h(u), h(v) \} } & \max \{ h(u), h(v) \} \le H_0 \\
    \end{cases},
\end{equation*}

where $0 < H_0 < H$ is the core's threshold. The probability $g(u, v)$ of an edge between two nodes with $\max \{ h(u), h(v) \} \le H_0$ (i.e. core--core edges)  is greater than the probability between two nodes whose heights satisfy $\min \{ h(u), h(v) \} \le H_0$ and $\max \{ h(u), h(v) \} > H_0$ (core--periphery edges), which is greater than the probability of the case that $\min \{ h(u), h(v) \} > H_0$ (periphery--periphery edges). Figure~\ref{fig:igam2} presents the adjacency matrix of a sampled IGAM2 network with parameters $c = 1.5, c_2 = 2.5, b = 3, H_0 = 2$, and $H = 6$.  

\amendRA{
We analyze the mathematical properties of IGAM2, which are similar to the properties of IGAM, in the Methods Section. Most of our proofs are based on a construction of a coupling of an IGAM2 network with two (simple) IGAM networks with parameters $(b, c_1, H)$ and $(b, c_2, H)$. The coupling is constructed such that the three graphs form an ordering based on the subgraph relation. 

\subsection{Directed and Continuous Versions}

The IGAM model has a natural \emph{directed extension}: for two nodes $u, v$ with heights $h(u)$ and $h(v)$ with $h(u) \le h(v)$ we create an edge from $u$ to $v$ with probability $\xi(u, v) = c^{-1-h(v)}$ and a directed edge from $v$ to $u$ with probability $\xi(v, u) = c^{-1-h(u)}$. This edge creation law  corresponds to the following philosophy: a non--famous node wants to connect to a prestigious node and a famous node does not want to connect to a non--famous one but it has better affinity for the nodes near its prestige $h(\cdot)$. This version of IGAM has also a sublinear dominating set, that is every node in the periphery has a directed edge to at least one node in the core w.h.p.. The proof of this fact is identical to the case of the simple model.

In the \emph{continuous} version of IGAM the height of a node $v$ is allowed to be any real number $h(v) \in [0, H]$ and the edge creation law remains the same as the simple version of IGAM. Moreover, similarly to \cite{tudisco2019nonlinear} the edge creation law $f(u, v)$ can be approximated by the limit as $\delta \to - \infty$ of a law $f_{\delta}(u, v)$ that involves the \emph{generalized mean} of $h(u)$ and $h(v)$, i.e. 

\begin{equation*} \tag{$\delta$--IGAM} \label{eq:delta_igam}
    f_\delta(u, v) = c^{-1- \left ( \frac {h^\delta(u) + h^\delta (v)} {2} \right )^{1 / \delta}}. 
\end{equation*}

The model given by \eqref{eq:delta_igam} can be treated as the \emph{scale--free} version of the logistic model of \cite{tudisco2019nonlinear} where the reverse ranking is replaced by the height function. The network has $n = b^H - 1$ nodes. If the heights $h$ are \emph{latent variables} drawn independently from a distribution with Cumulative Density Function (CDF) equal to 

\begin{equation*}
    \Pr [h(v) \le t] = \begin{cases}
        0 & t < 0 \\
        \frac {b^t - 1} {b^H - 1} & t \in [0, H] \\
        1 & t > H
    \end{cases},
\end{equation*}

then we can easily show that the continuous model has a sublinear dominating set by partitioning $[0, H]$ to intervals of the form $[t_i, t_i + \Delta t]$ and generalizing the analysis of the discrete model as $\Delta t$ becomes infinitesimal.

Finally, a mathematical and empirical study and a comparison between the extensions of IGAM and logistic core--periphery models are interesting questions to be addressed in future work, and lie beyond the scope and length of the current paper. 

}

\section{Conclusions}

The present paper observes a connection between the core--periphery structure of networks and dominating sets. We devise a simple generative model which facilitates this connection and fit it to real--world data validating our observations. We believe it is worthwhile to explore the algorithmic implications of this connection further.

\section*{Methods}

\amendRA{
\subsection*{Reproducibility}

Code and data needed to exactly reproduce are provided in the form of a Jupyter notebook and is available here\cite{source_code}. The software has been developed in Python by the author and uses the following open--source libraries: numpy \cite{numpy}, scipy \cite{scipy}, networkx \cite{networkx}, matplotlib \cite{matplotlib}, pandas \cite{pandas}, and seaborn\cite{seaborn}.

\subsection*{Qualitative Results Addendum}

The analytical results which are briefly presented in \cref{sec:qualitative} can be found below for the first three levels of the hierarchy. Groups enclosed in parentheses correspond to separate levels. In the faculty hiring networks the ``All others'' node represents all non--US institutions:

\begin{itemize}
    \item \emph{world--trade}: (Finland) (Hungary, Slovenia, Singapore, Chile) (Salvador, Iceland, Kuwait, Rep., Belgium, Poland, Moldava., Austria, Germany, Indonesia, Guatemala, Bolivia, Paraguay, Australia, Africa, Of)
    \item \emph{london--underground}: (Bank) (Baker Street, Canning Town) (Kings Cross St. Pancras, Stratford, Willesden Junction, Earls Court)
    \item \emph{open--arilines}: (AMS) (FRA, CDG) (IST, MUC, ATL, PEK)
    \item \emph{cs--faculty}: (All others) (University of Illinois, Urbana Champaign, MIT) (Purdue University, University of Texas, Austin, Carnegie Mellon University, Stanford University)
    \item \emph{history--faculty}: (All others) (Harvard University, Yale University, University of Chicago, University of Wisconsin, Madison, Columbia University) (UC Berkeley, UCLA, Princeton University, University of Michigan, University of Pennsylvania, Stanford University, Johns Hopkins University, Rutgers University, University of Virginia, Cornell University, University of Texas, Austin, New York University, Indiana University, Northwestern University, Ohio State University, University of Illinois, Urbana Champaign, University of North Carolina, Chapel Hill, Duke University, Brown University, University of Minnesota, Minneapolis, Michigan State University, UC San Diego, UC Santa Barbara, Brandeis University, University of Washington)
    \item \emph{business--faculty}: (All others) (University of Michigan, University of Texas, Austin) (Ohio State University, Indiana University, Pennsylvania State University, University of Pennsylvania)
\end{itemize}

\subsection*{Global Clustering Coefficient of IGAM}

For the number of closed triplets (i.e. triangles) we have

    \begin{equation} \label{eq:closed_triplets}
    \begin{split}
        \mathbb E[T_C] & = \sum_{w : h(w) = h(v) + 1}^H \sum_{v: h(v) = h(u) + 1}^{h(w)} \sum_{u: h(u) = 0}^{h(v)} b^{h(u) + h(v) + h(w)} c^{-3-2h(u) - h(v)} \\
        & = \frac {1} {c^3} \sum_{w : h(w) = h(v) + 1}^H \sum_{v : h(v) = h(u) + 1}^{h(w)} b^{h(v) + h(w)} c^{-h(v)} \sum_{h(u) = 0}^{h(v)} b^{h(u)} c_2^{-2h(u)} \\
        & = \frac {1} {c^3} \sum_{w : h(w) = h(v) + 1}^H \sum_{v : h(v) = 0}^{h(w)} b^{h(v) + h(w)} c^{-h(v)} \Theta \left ( \frac {b^{h(v)}} {c^{2h(v)}} \right ) \\
        & = \frac {1} {c^3} \sum_{w : h(w) = h(v) + 1}^H \sum_{v : h(v) = 0}^{h(w)} b^{h(w)} b^{h(w)} \Theta \left ( \frac {b^{2h(w)}} {c^{3h(w)}} \right ) = \Theta \left ( \frac {b^{3H}} {c^{3H + 3}} \right ).
    \end{split}
    \end{equation}
    
    For the number of open triplets we have that
    
    \begin{equation} \label{eq:open_triplets}
    \begin{split}    
        \mathbb E[T_R] & = \sum_{w : h(w) = h(v) + 1}^H \sum_{v: h(v) = h(u) + 1}^{h(w)} \sum_{u: h(u) = 0}^{h(v)} b^{h(u) + h(v) + h(w)} \gamma_{uvw} \\
        & \le 3 \sum_{w : h(w) = h(v) + 1}^H \sum_{v: h(v) = h(u) + 1}^{h(w)} \sum_{u: h(u) = 0}^{h(v)} b^{h(u) + h(v) + h(w)} c^{-2-2h(u)} \\
        & = \frac {3} {c^2} \sum_{w : h(w) = h(v) + 1}^H \sum_{v: h(v) = 0}^{h(w)} b^{h(v) + h(w)} \Theta \left ( \frac {b^{h(v)}} {c^{2h(v)}}\right ) \\
        & = \frac {3} {c^2} \sum_{w : h(w) = 0}^H b^{h(w)} \Theta \left ( \frac {b^{2h(w)}} {c^{2h(w)}}\right ) = \Theta \left ( \frac {b^{3H}} {c^{2H + 2}} \right ).
    \end{split}    
    \end{equation}
    
    Similarly, $\mathbb E[T_R] \ge 3 \sum_{w : h(w) = h(v) + 1}^H \sum_{v: h(v) = h(u) + 1}^{h(w)} \sum_{u: h(u) = 0}^{h(v)} b^{h(u) + h(v) + h(w)} c^{-2-2h(v)} = \Theta \left ( \frac {b^{3H}} {c^{2H + 2}} \right )$. Therefore, $\mathbb E[T_R] = \Theta \left ( \frac {b^{3H}} {c^{2H + 2}} \right )$. 

\subsection*{Properties of IGAM2} 

We describe the mathematical properties of IGAM2. First, we construct a coupling between IGAM2 and IGAM which we can use a proxy for the behaviour of IGAM2. 

\paragraph{Remark.} Throughout the proofs we use the following remark: For every two positive integers $s, t$ with $s < t$ and for a positive integer constant $b \ge 2$ we have that $(1 - 1 / b) b^t \le b^t - b^{t - 1} \le b^t - b^s \le b^t$. Therefore $b^t - b^s = \Theta (b^t)$ with constants $C_1 = 1 - 1 / b$ and $C_2 = 1$. 

\paragraph{Coupling Construction.} We consider a randomly generated network $G \sim \mathrm {IGAM2}(b, c_1, c_2, H_0, H) \equiv \mathrm{IGAM}(b, c_1, H)$ for $1 < c_1 \le c_2 < b$ and $0 \le H_0 \le H$ with edge law $g$. We also consider a network $G' \sim \mathrm{IGAM2}(b, c_2, c_2, 0, H) \equiv \mathrm{IGAM}(b, c_2, H)$ and a network $G'' \sim \mathrm{IGAM2}(b, c_1, c_1, 0, H)$ with edges law $g'$ and $g''$ coupled with $G$ as follows: 

\begin{itemize}
    \item $\Pr \left [ (u, v) \in E(G) | (u, v) \in E(G') \right ] = 1$ and $\Pr \left [ (u, v) \in E(G) | (u, v) \notin E(G') \right ] = \tfrac {g(u, v) - g'(u, v)}{1 - g'(u, v)} \in [0, 1]$.
    \item $\Pr \left [ (u, v) \in E(G'') | (u, v) \in E(G) \right ] = 1$ and $\Pr \left [ (u, v) \in E(G'') | (u, v) \notin E(G) \right ] = \tfrac {g''(u, v) - g(u, v)} {1 - g(u, v)} \in [0, 1]$.  
\end{itemize}

Under this coupling, which we denote as $\nu$, we have that $\Pr \left [ (u, v) \in E(G) \right ]  = \Pr \left [ (u, v) \in E(G) \right | (u, v) \in E(G') ] \Pr [(u, v) \in E(G')] + \Pr \left [(u, v) \in E(G) | (u, v) \notin E(G') \right ] \Pr [(u, v) \notin E(G') ]  = g'(u, v) \cdot 1 + (1 - g'(u, v)) \cdot \tfrac {g(u, v) - g'(u, v)} {1 - g'(u, v)} = g'(u, v) + g(u, v) - g'(u, v) = g(u, v)$ and, similarly, $\Pr \left [ (u, v) \in E(G'') \right ] =   g''(u, v)$. The coupling also satisfies that $G'$ is a subgraph of $G$ ($G' \subseteq G$) since $(u, v) \in E(G')$ implies $(u, v) \in E(G)$. Moreover, $G$ is a subgraph of $G''$ ($G \subseteq G''$) since every edge of $G$ belongs to the edge set of $G''$. 
    
\paragraph{Sublinear Dominating Set.} We let $(G', G, G'') \sim \nu$. We know that $G'$ is generated from a simple IGAM model therefore it has a dominating set of size $b^{O(\log (2 c_2 H \log b) / \log (b / c)} = b^{o(H)} = o(n)$. Since $G' \subseteq G$, the dominating set of $G$ has size at most the dominating set of $G'$. Therefore, $G$ has a dominating set of size $b^{O(\log (2 c_2 H \log b) / \log (b / c)} = b^{o(H)} = o(n)$. 

\paragraph{Degree Distribution.} We fix a node $u \in V$. We have the following

\begin{enumerate}
    \item If $h(u) > H_0$ then the law that is obeyed is $g(u, v) = c_2^{-1-\min \{h(u), h(v) \}}$. From the simple IGAM model we have calculated the degree in this case to be $\Theta \left (b^{H + 1} / c_2^{h(u) + 1} \right )$. 
    \item If $h(u) \le H_0$ then 
    \begin{equation*}
    \begin{split}
        \bar d_h & \approx \sum_{r = 0}^{H_0} b^r c_1^{-\min \{ h(u), r \} - 1} + \sum_{r = H_0 + 1}^H b^r c_2^{-\min {h(u), r \} - 1}} & = \Theta \left ( \frac {b^{H_0 + 1}} {c_1^{h(u) + 1}} \right ) + \sum_{r = H_0 + 1}^H b^r c_2^{-1-h(u)} \\ & = \Theta \left ( \frac {b^{H_0 + 1}} {c_1^{h(u) + 1}} \right ) + \Theta \left ( \frac {b^{H + 1}} {c_2^{h(u) + 1}} \right ).
    \end{split}
    \end{equation*}
\end{enumerate}

    Therefore, every node, parametrized by its height $h$ has average degree 
    
    \begin{equation*}
        \bar d_h \approx \begin{cases} 
            \Theta \left (\frac {b^{H + 1}} {c_2^{h + 1}} \right ) & h > H_0, \\
            \Theta \left ( \frac {b^{H_0 + 1}} {c_1^{h + 1}} \right ) + \Theta \left ( \frac {b^{H + 1}} {c_2^{h + 1}} \right ) & h \le H_0
        \end{cases}.
    \end{equation*}

    To bound the average number of edges we refer to the coupling $\nu$ and deduce that the average number of edges $\bar m$ of $G$ is at most the average number of edges of $G'$, say $\bar m' = \Theta (b^{2H} / c_2^H)$ (as we showed in the main part of the paper) due to the subgraph relationship. Therefore, the average number of edges is $\bar m = O(b^{2H} / c_2^H)$. A better bound can be obtained by calculating the expected value analytically using the form of $\bar d_h$ we derived above. Namely,
    
    \begin{equation*}
        \begin{split}
            \bar m & = \sum_{h = 0}^{H_0} b^h \bar d_h + \sum_{h = H_0 + 1}^H b^h \bar d_h \\
            & = \sum_{h = 0}^{H_0} b^h \left [ \Theta \left ( \frac {b^{H_0 + 1}} {c_1^{h + 1}} \right ) + \Theta \left ( \frac {b^{H + 1}} {c_2^{h + 1}} \right ) \right ] + \sum_{h = H_0 + 1}^H b^h \Theta \left (\frac {b^{H + 1}} {c_2^{h + 1}} \right )  \\
            & = \Theta \left ( \frac {b^{2H_0}} {c_1^{H_0}} \right )  +  \Theta \left ( \frac {b^{H + H_0}} {c_1^{H_0}} å\right ) + \Theta \left ( \frac {b^{2H}} {c_2^H} \right ). 
        \end{split}
    \end{equation*}
    
    We still observe that $\bar m = O(b^{2H} / c_2^H)$. Moreover, using the fact that the edges $\bar m''$ of $G''$ are $\Theta (b^{2H} / c_1^H)$ we get, in the same logic, that $\bar m \ge \bar m''$, and, thus $\bar m = \Omega (b^{2H} / c_1^H)$. Note that setting $c_1 = c_2$ and $H_0 = 0$ recovers the result for the simple IGAM model.
    
    \paragraph{Small-world Behaviour.} We let $(G', G, G'') \sim \nu$. Since $G' \subseteq G$, the diameter of $G$ is at most the diameter of $G'$ because every path between two nodes in $G'$ is a path in $G$. Since the diameter of $G'$ is close to $\Theta (\log b / \log (b / c_2)) = O(1)$ a.a.s., then the diameter of $G$ is also close to $O(1)$ a.a.s..
    
    \paragraph{Global Clustering Coefficient.} Let $(G', G, G'') \sim \nu$. Let $uvw$ be a triplet in $G$ such that $h(u) \le h(v) \le h(w)$. The probability that $uvw$ is a triangle in $G'$ is $\beta_{uvw}'$, $\beta_{uvw}$ if $uvw$ is a triangle in $G$ and $\beta_{uvw}''$ if $uvw$ is a triangle in $G''$. From the subgraph relationship we have that $c_2^{-3-2h(u)-h(v)} = \beta_{uvw}' \le \beta_{uvw} \le \beta_{uvw}'' = c_1^{-3-2h(u) - h(v)}$. Therefore, the number of triangles $T_C$ (respectively $T_C'$ for $G'$ and $T_C''$ for $G''$) satisfies $\mathbb E[T_C'] \le \mathbb E[T_C] \le \mathbb E[T_C'']$. Using \cref{eq:closed_triplets} we deduce that $\mathbb E[T_C''] = \Theta \left ( \frac {b^{3H}} {c_1^{3H + 3}} \right )$  and $\mathbb E[T_C'] = \Theta (b^{3H} / c_2^{3H + 3})$. 
    
    The probability $\gamma_{uvw}$ of $uvw$ being a triplet in $G$ (respectively $\gamma_{uvw}'$ in $G'$ and $\gamma_{uvw}''$ in $G''$) satisfies $3c_2^{-2-2h(v)} \le \gamma_{uvw}' \le \gamma_{uvw} \le \gamma_{uvw}'' \le 3c_2^{-2-2h(u)}$. The expected number of triplets is denoted by $\mathbb E [T_R]$ ($\mathbb E[T_R']$ for $G'$ and $\mathbb E[T_{R}'']$ for $G''$) can be found by using \cref{eq:open_triplets}. If we execute the sum mutatis mutandis, we arrive at the fact that $\Omega (b^{3H} / c_2^{2H + 2} = \mathbb E[T_R] = O (b^{3H} / c_1^{2H + 2})$. McDiarmid's Inequality\cite{doob1940regularity} states that $\Pr \left [ T_C \le \mathbb E[T_C] + O(b^H) \right ] = 1 - O \left ( e^{-b^H} \right )$, and $
    \Pr \left [ T_R \ge \mathbb E[T_R] - O(b^H) \right ] = 1 - O \left ( e^{-b^H} \right )$.
    because $T_C, T_R$ are $\Theta(b^H)$--Lipschitz functions [In general, for a graph $G$ with $n$ nodes the number of triangles of $G$ as a function of the edge variables is a $3n$--Lipschitz per edge, since deleting or adding an edge can change the number of triangles by $3n$, and, similarly, the number of triplets is a $2n$--Lipschitz function since each edge is part of at most $2n$ paths on 3 vertices]. Thus, with probability $1 - O \left ( e^{-b^H} \right )$ we have that $\tfrac {T_C} {T_R} \le \tfrac {\mathbb E [T_C]} {\mathbb E [T_R]} + O (b^{-H}) = O \left ( \tfrac {c_2^{2H + 2}} {c_1^{3H + 3}} + b^{-H} \right )$. 
    \paragraph{Core--periphery Conductance.} Let $(G', G, G'') \sim \nu$. Let the partition $(S_\tau, \bar S_\tau)$ be at level $\tau$, i.e. all nodes with height $h \le \tau$ and the periphery $\bar S$ with $h \ge \tau$. From the subgraph relationship we get that $e'(S_{\tau}, \bar S_{\tau}) \le e(S_{\tau}, \bar S_{\tau}) \le e''(S_{\tau}, \bar S_{\tau})$, and subsequently $\mathbb E [e'(S_\tau, \bar S_{\tau})] \le \mathbb E[e(S_\tau, \bar S_{\tau}) \le \mathbb E [ e''(S_\tau, \bar S_{\tau})]$.  Thus $\bar \phi'(S_\tau) \le \bar \phi(S_\tau) \le \bar \phi''(S_\tau)$. Using the fact about the core periphery conductance we proved for the simple IGAM model, since $G', G''$ are equivalently produced from the simple IGAM model, we get that, on expectation, $\Omega \left ( \tfrac {b^H} {c_2^\tau} \right ) = \bar \phi(S_{\tau}) = O \left ( \tfrac {b^H} {c_1^\tau} \right )$. If we take $\tau = H_0 = O(\log H)$ to be the core, we can deduce that the core conductance is $\Theta (b^H / H)$ as in the case of the simple IGAM model.

\subsection*{Data Preprocessing}

We have ignored directionality in the examined networks and have removed nodes with degree less than or equal to 4 (except in the london--underground network where almost all degrees are very small). The removal of nodes with degree less than or equal to 4 is done (i) to remove outlier nodes and, (ii) to refer to the removal of non--engaged nodes. 
}

\bibliography{references}



\section*{Acknowledgements}

The author would like to thank Jon Kleinberg for his useful suggestions during the preparation the present manuscript. The author would like to thank the anonymous referees for their valuable feedback. Supported in part by a Vannevar Bush Faculty Fellowship. The article is openly available at \url{www.nature.com/articles/s41598-021-94105-8}

\section*{Author contributions statement}

M.P. is the sole author of the paper, and undertook the entirety of the research. 

\section*{Additional information}



\subsection*{Accession Codes}

The data used in this study are publicly available and are located in the following resources

\begin{itemize}
    \item \href{http://casos.cs.cmu.edu/computational_tools/datasets/external/world_trade/index2.html}{world--trade}~\cite{de2018exploratory}.
    \item \href{http://advances.sciencemag.org/cgi/content/full/1/1/e1400005/DC1}{\{cs, history, business\}--faculty}~\cite{clauset2015systematic}.
    \item \href{http://networkrepository.com/polblogs.php}{polblogs}~\cite{adamic2005political}.
    \item \href{https://toreopsahl.com/datasets/#usairports}{airports}~\cite{colizza2007reaction}.
    \item \href{https://www.cs.cornell.edu/~arb/data/spatial-OpenFlights/}{open--airlines}~\cite{jia2019random}.
    \item \href{https://www.cs.cornell.edu/~arb/data/spatial-Celegans/}{celegans}~\cite{Kaiser-2006-placement}.
    \item \href{https://www.cs.cornell.edu/~arb/data/spatial-underground-London/}{london--underground}~\cite{jia2019random}.
\end{itemize}

The Jupyter notebook for reproducing the results of this study is also available in \cite{source_code}. 

\subsection*{Competing Interests}

The author declares that there are no competing interests for this manuscript.

\subsection*{Ethical Statement}

The current paper proposes a theoretical model, its properties and fits it to real--world data. Thus, there are no ethical concerns.

\newpage

\begin{figure}[ht] 
\centering

\fbox{\begin{minipage}{0.99\textwidth}

\amendRA{
\begin{enumerate}
\item
  The input is provided as a dataset of \(m\) edges
  \(\mathcal D = \{ e_1, \dots, e_m \}\).
\item
  Calculate the degree \(\bar y_u\) of every node \(u\) in the sample.
\item
  We sort the degrees in descending order.
\item
  For all fanouts \(b \in \{ 2, \dots, n - 1 \}\)
\item
  We build a tree by attributing heights to the nodes in descending
  order of their degree
\item
  We calculate
  \(\bar z_h = \log \left ( \sum_{u: h(u) = h} \bar y_u \right )\), that
  is the log-total number of edges on level \(h\) as indicated by the
  samples.
\item
  We fit a linear least squares relation between \(h\) and \(\bar z_h\)
  that has the form \(\hat z_h = a h + b\)
\item
  We calculate \(c = b \cdot e^{-a}\), since the slope \(a\) is roughly
  \(\log (b / c)\).\\
\item
  We calculate the log--likelihood of the parametrization which equals $$\sum_{u < v} \left ( \mathbf 1 \{(u, v) \in \mathcal D \} \log ( c^{-1 - \min \{ h(u), h(v) \}}) + (1 - \mathbf 1 \{ (u, v) \in \mathcal D \}) \log (1 - c^{-1 - \min \{ h(u), h(v) \}} ) \right ). $$

\item
  We return the set of parameters that maximize the computed likelihood.

\item
  \emph{(Optional: Swaps)} Iterate on every edge $(u, v) \in \mathcal D$ and swap $h(u)$ with $h(v)$ if the log--likelihood increases, otherwise do nothing. Iterate until no more swaps are possible.

\end{enumerate}
}
\end{minipage}}

\captionof{algorithm}{IGAM Fitting Algorithm Pseudocode.}
\label{alg:igam_fit}

\end{figure}

\end{document}